\documentclass{article} 
\usepackage{geometry}
\usepackage{graphicx} 
\usepackage{float}
\usepackage[round]{natbib}
\usepackage[colorlinks=true, urlcolor=blue, linkcolor=blue, citecolor=blue]{hyperref}
\usepackage{booktabs}

\setcitestyle{authoryear,open={[},close={]},semicolon}
\title{The Statistical Validation of Innovation Lens} 
\author{
    Giacomo Radaelli \\  
    Innovation Lens \\
    \url{gradaelli@gradcenter.cuny.edu}
    \and
    Jonah Lynch \\
    Innovation Lens \\
    \url{jonah@innovationlens.org}
}
\date{19 August 2025}

\begin{document}

\maketitle

\begin{abstract}
    Information overload and the rapid pace of scientific advancement make it increasingly difficult to evaluate and allocate resources to new research proposals. Is there a structure to scientific discovery that could inform such decisions? We present statistical evidence for such structure, by training a classifier that successfully predicts high-citation research papers between 2010-2024 in the Computer Science, Physics, and PubMed domains. 
\end{abstract}

\section{Introduction}

The exponential growth in the quantity of scientific publications necessitates new tools to manage information overload. The philosophy of science has debated what mechanism, if any, drives scientific progress.\citep{popp-logi,lakatos1976proofs,kuhn2012structure} Metascience\citep{ioannidis_why_2005,salatino2020ontologyextractionusagescholarly} is an updated, often more quantitative study of the same problem\citep{rubin_replication_2025}, which has recently gained important tooling from language models, text vectorization, and graph representation of author relationships and citation networks.\footnote{See Litmaps, Scite, Connected Papers, etc.}\citep{NBERw33821,2023} Some tools to represent scientific discourse using spatial metaphors have also been developed using dimension reduction methods like Uniform Manifold Approximation and Projection (UMAP).\citep{mcinnes2020umapuniformmanifoldapproximation}\footnote{See Nomic Atlas for example.} Using Large Language Models (LLMs), several companies have developed automated tools to produce literature reviews\footnote{Elicit, Potato, etc.}, and some even work toward the automation of the entire scientific process.\citep{lu2024aiscientistfullyautomated} It is a short step from studying the structure of scientific progress to attempting to predict it. Some researchers have tried to find correlations between articles about science on Wikipedia, as represented by the hyperlinks between articles, and the future award of the Nobel Prize to these discoveries.\citep{ju2020networkstructurescientificrevolutions} 

We wondered if the millions of scientific articles on public repositories contain hidden patterns. While it is impossible for a human to exhaustively review more than a few thousand articles, perhaps computational techniques could make much larger collections amenable to quantitative analysis. Could an overall structure be found in unstructured text, even with all the noise inherent in a public, non-peer-reviewed repository like arXiv or PubMed? If so, it might open a road toward solving the problem of information overload.

\section{Method}
We trained a classifier to identify the topics of future articles that were likely to receive a high number of citations in the 24 months following publication. We reasoned that, while a highly imperfect measure, high citation count over a short period of time is the best single metric available to indicate the relative importance of a scientific paper. If indeed our classifier could predict which papers would receive more attention and citations, it would be a useful tool in directing the limited attention resources of scientists and funding agencies toward the topics that are most likely to prove valuable.

In order to normalize the natural growth in citation count as an article ages, we identified targets as the $top-P$ percentile articles. The classifier performs optimally when predicting the top 15\% of articles, that is, the set of articles that received more citations than 85\% of the articles published during the same month.\footnote{Citation counts were accessed via API from \url{openalex.org}. Because of the nature of preprint repositories and the limitations of OpenAlex, not all articles used in our classifier have citation data available. This limitation presumably skews our measurement of the performance of both the algorithm and baseline by the same amount.}

Our classifier was built by encoding more than 30 million scientific articles as vectors in a high-dimensional space and using various measures of that space to predict where future high-citation research would occur. No prompting or generative AI is involved, but we do employ an LLM for text vectorization. We tried three different language models to verify that our results do not depend on using a recent model, which might overfit because it already contains the data we are trying to predict. All three language models gave similar results, so we do not detail this experiment here beyond noting that open-source models trained five years ago give results comparable to those obtained from very recent proprietary models.\footnote{There are deep reasons to expect this result.\citep{huh2024platonicrepresentationhypothesis,jha2025harnessinguniversalgeometryembeddings}}

In the high-dimensional latent space of the collection of articles on arXiv and PubMed, which contains information about nearly all recent developments in the fields of physics, computer science, and medicine, our classifier identifies the coordinates of locations where future high-citation articles are likely to be published. The value of these coordinates was validated by back-testing each month from 2010 through 2024, using all articles up until the cutoff month to predict targets in the subsequent 24-month period. Each run of the algorithm on a collection of $N_{articles}$ generates $N_{predictions}$, where $N_{articles}$ is several orders of magnitude larger than $N_{predictions}$.\footnote{To be more precise, $N_{articles}$ is about 100 times larger than $N_{predictions}$ for Computer Science articles on arXiv, and about 1000 times larger for Physics on arXiv and for PubMed.}

As a baseline for comparison, we made a model of incremental scientific research. Typically, researchers belong to a tradition of inquiry and either continue within that vein, or step outside their tradition by a small amount in subsequent research projects. We operationalized this approach by using the coordinates of the existing scientific research distribution and sampling from it to produce a distribution of the locations of $N_{predictions}$ potentially valuable new articles. During validation, these baseline predictions were treated in the same way as algorithmic predictions.\footnote{We attempted to improve this baseline by sampling only from the subset of articles that were in the $top-P$ percentile by citation count, to model the action of trend-following researchers who choose to focus their work on fields where highly cited articles are already present. However, we judge this baseline to be invalid for the purposes of validation, because it introduces potentially significant biases. First of all, a symmetric filter cannot be applied to the algorithmic predictions. Even more troubling, it is likely that our model of "trend-following researchers" suffers from future leak because the citation counts in the dataset are those measured at test-time (August 2025), not at the past time the supposed trend-followers would have had to make their choice. We have no way to replicate in backtesting the true information landscape in which they would have found themselves years ago. Finally, as mentioned, we have citation data only for a fraction of the total articles available. (For the record, even with all these sources of bias, our algorithm still outperforms the modified baseline.)} 

For each model, we calculate the true and false positive and negative values ($TP$, $FP$, $FN$, and $TN$) and related statistical measures like MCC, F1, precision, and recall, by comparing the location of a prediction to the locations of articles in the test set (the subsequent 24 months) within a radius $\epsilon$ of the prediction.\footnote{The parameter $\epsilon$ can be interpreted as a measure of the specificity of a prediction: if it is small, predictions can be precisely characterized as regarding a specific topic. Larger $\epsilon$ can still be a useful signal, by indicating a more general topic of interest.} Articles belonging to the $top-P$ percentile of the test set, by citation count, are considered targets. 

Naively, one might expect to measure classifier performance using a simple comparison between predictions and targets (a target is a TP if within $\epsilon$ of a prediction; an article that is not a target is a FP if within $\epsilon$ of a prediction; a target is a FN if it is not within $\epsilon$ of a prediction; an article that is not a target is a TN if it is not within $\epsilon$ of a prediction). While clean and easy to understand, this approach would have the disadvantage of not adequately considering the cluster of articles that follow a major breakthrough, and which indicate the importance of a breakthrough (and therefore, the value of a prediction) by their number, even though the majority of them will never reach the top-15\% percentile of citations. We therefore built a more sophisticated method to calculate the TPR and FPR of our models.

An article that is not a target is counted as a TP if it occurs within $\epsilon$ of a target, and chronologically after that target's publication date.\footnote{Throughout, we consider the publication date to be the first upload date associated with a preprint. Many articles undergo important revisions during the peer-review process that leads to traditional publication, but we consider that the main predictive signal should be associated with the first public expression of the ideas in the article, not with its final form months or years later.} 

\begin{figure}[H]
    \centering
    \includegraphics[width=0.6\textwidth]{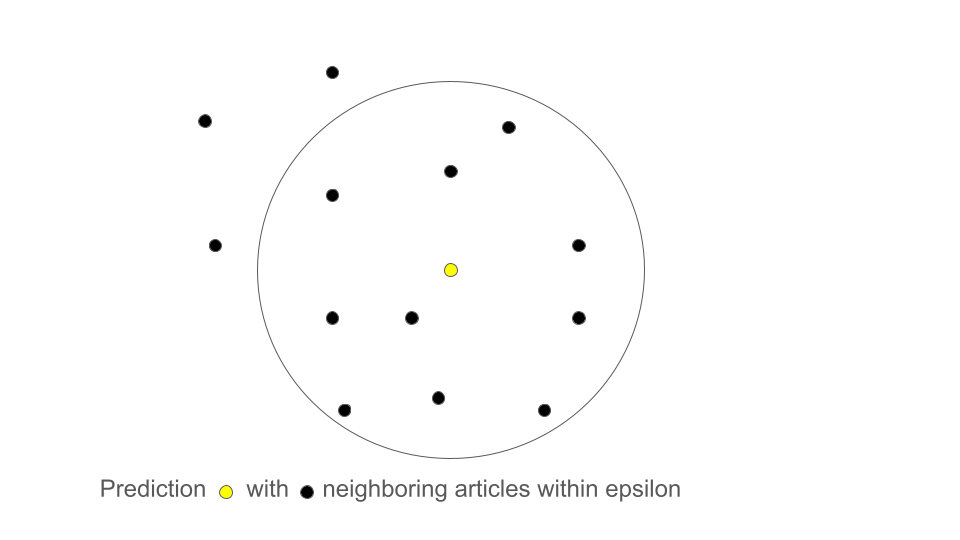}
    \caption{The classifier produces $N_{predictions}$: one shown, in gold. The test set contains $X$ targets which are the $top-P\%$ most cited articles within the test set.}
    \label{fig:validation1}
\end{figure} 

\begin{figure}[H]
    \centering
    \includegraphics[width=0.6\textwidth]{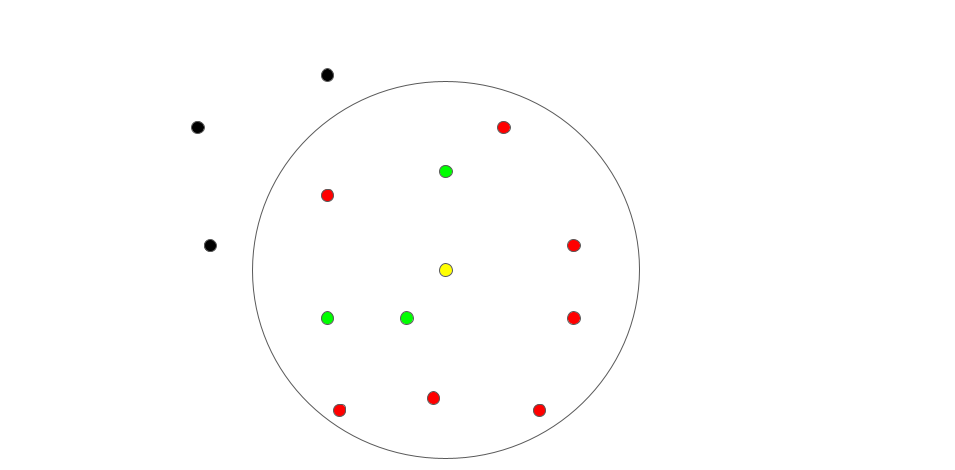}
    \caption{Articles within radius $\epsilon$ of a prediction (gold) are classified TP if they have $top-P\%$ citations. Other articles within $\epsilon$ of a prediction, and chronologically posterior to a TP also within $\epsilon$ of the same prediction with $top-P\%$ citations, are also classified as TP (green). All other articles are classified as negatives, with TN/FN established as described in the main text.}
    \label{fig:validation2}
\end{figure}

An article is a FN if it is not within $\epsilon$ of a prediction but does have $top-P\%$ citations. Symmetrically with the scoring of FP, all chronologically subsequent articles in the test set within $\epsilon$ of an unpredicted $top-P\%$ target are also counted as FNs. Finally, articles that have neither $top-P\%$ citations nor are within $\epsilon$ of a prediction are classified as TN. This method has the advantage that it represents clusters of articles on the same topic more faithfully than the first method, which treats each article independently of its context. 

Interestingly, the clusters around successful predictions are about an order of magnitude larger than the clusters around unpredicted targets.\footnote{This value is inferred by comparing the values of TP and FN in the validation method described, to the results that would be achieved using a naive method of counting TP as only the single articles with higher than $top-P$ citation count. Using our chosen validation method, TP increases by approximately a factor of 10 faster than the rate of FN increase.} This may mean that our classifier's predictions are indeed capturing a robust structure of the latent space: the value of the predicted targets (TP), as inferred by the quantity of follow-on articles on the same topic, is in general higher than the value of unpredicted targets (FN).

\section{Results}
  
First, we measured our classifier's performance on the Computer Science section of the arXiv repository. From the values of $TP$, $FP$, $FN$, and $TN$, we calculate the True Positive Rate and the False Positive Rate as

\[
TPR = \frac{TP}{TP + FN} \quad  \quad FPR = \frac{FP}{FP + TN}.
\]

\begin{figure}
    \centering
    \includegraphics[width=0.6\textwidth]{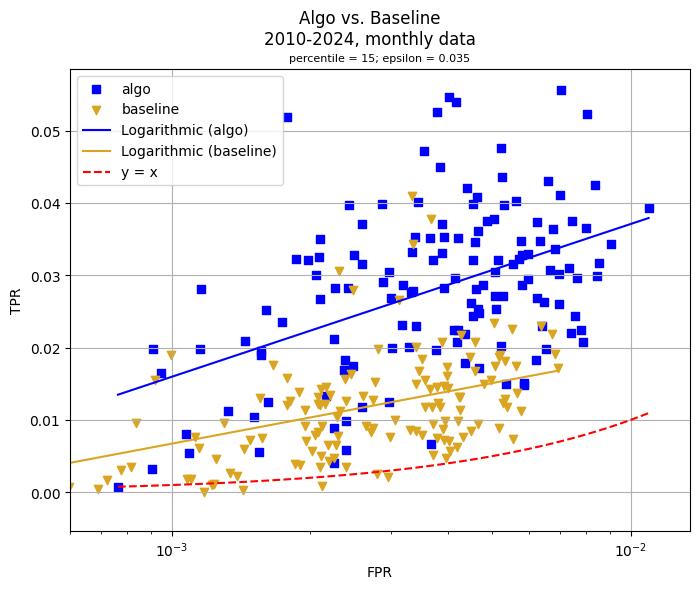}
    \caption{Classifier performance on arXiv (Computer Science) from 2010 to 2024 with parameters $\epsilon = 0.035$ and top-15\% as targets, with logarithmic fits.}
    \label{fig:cs2010-2024}
\end{figure}

We performed a grid search over parameters: the $top-P$ percentiles between 1 and 20, a range of values of $\epsilon$, and the month-by-month data from 2010 to 2024. Selecting the top 15\% of citation count and setting $\epsilon = 0.035$, we can summarize performance over the last 15 years in a single chart in Figure~\ref{fig:cs2010-2024}. This chart contains a $y=x$ line (in red) to represent purely random classifier performance, the baseline model which represents how scientists traditionally choose topics for future research (in yellow), and our algorithm (in blue). The x-axis is represented here on a log scale.

There is broad variance in the signal, but there is also a clear signal. The logarithmic fitted line is meant only as a rule-of-thumb measure that permits us to assert that over a statistically significant sample, the algorithm can be expected to perform twice as well as the baseline. 

\subsection{Results for Other Repositories}

\begin{figure}
    \centering
    \includegraphics[width=0.6\textwidth]{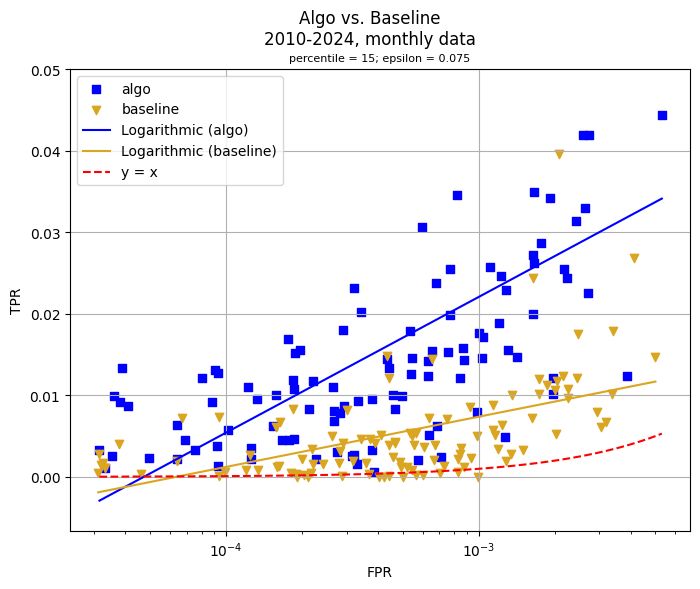}
    \caption{Classifier performance on arXiv (Physics) from 2010 to 2024 with parameters $\epsilon = 0.075$ and top-15\% as targets, with logarithmic fits.}
    \label{fig:physics2010-2024} 
\end{figure}

We then applied the classifier to Physics articles on arXiv and produced the chart in Figure~\ref{fig:physics2010-2024}. On this domain the classifier approximately triples baseline performance, and we are still optimizing the parameters!

We also applied the classifier to the Mathematics domain on arXiv. Interestingly, we were unable to find parameters that gave better-than-baseline performance. We interpret this result to mean that different subjects present different distributions and internal structures, and only some of them are amenable to the method used by our classifier.\footnote{We did not try to apply the classifier to other domains on arXiv, like biology, quantitative finance, and electrical engineering, because they contain a much smaller dataset than is present for CS and Physics. In the future, we plan to repeat the experiment with dedicated repositories like bioRxiv.} 

\begin{figure}
    \centering
    \includegraphics[width=0.6\textwidth]{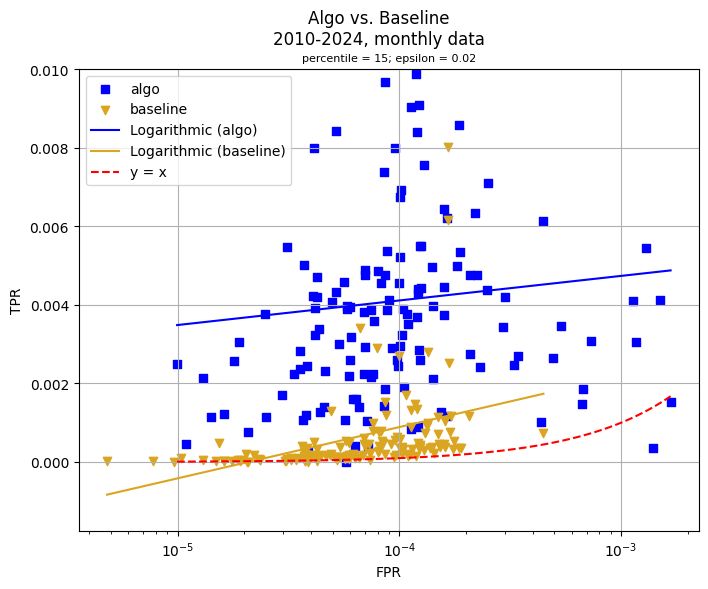}
    \caption{Classifier performance on PubMed from 2010 to 2024 with parameters $\epsilon = 0.02$ and top-15\% as targets, with a logarithmic fit.}
    \label{fig:pubmed2010-2024}  
\end{figure}

Finally, we moved to the PubMed repository. It contains an order of magnitude more articles than arXiv, spread across many subdisciplines of medicine. This increase in article quantity causes an increase in density in the normalized latent space of our models, so parameters were correspondingly reduced. Using $top-P = 15\%$ and setting $\epsilon$ to $0.02$, we produced the results in Figure~\ref{fig:pubmed2010-2024}.

These results are so much better than baseline that they seemed suspicious, so we extensively validated them across many parameters. We have been unable to find an error in our calculations, so we publish them here. We found that performance is highly dependent on the choice of $\epsilon$, and quickly collapses for values above $0.05$. This is encouraging: it implies that the algorithm performs best at high specificity (recall that because $\epsilon$ is a radius in latent space, the smaller the value the more specific the prediction.) We also note the small values for TPR and FPR, which are about an order of magnitude smaller than the similar calculations for the data from arXiv above. This is due to two features of the data in our PubMed analysis: first, the number of predictions we produce for each month in the PubMed data is actually smaller than the number of predictions for a month of arXiv data, which reduces the possible TP and FP values and hence the numerator of the FPR and TPR calculations.\footnote{Due to the size of the PubMed database, we are currently applying some restrictions to the algorithm in order to reduce computational complexity. This has the effect of reducing the overall number of predictions generated.} Secondly, since the dataset is so much bigger in PubMed, the FN and TN values (and the denominator of the FPR and TPR calculations) are correspondingly much larger.\footnote{A typical month's predictions for PubMed take $\approx60,000$ articles as context data and produce $\approx~20$ predictions. Even if every prediction correctly classifies multiple targets, $TP$ will be orders of magnitude smaller than $FN$, which will be (by definition) some fairly large fraction of $\approx60,000$, which will produce a very small number for $TPR$. The same logic holds for $FPR$.}

\begin{figure} 
    \centering
    \includegraphics[width=1.0\textwidth]{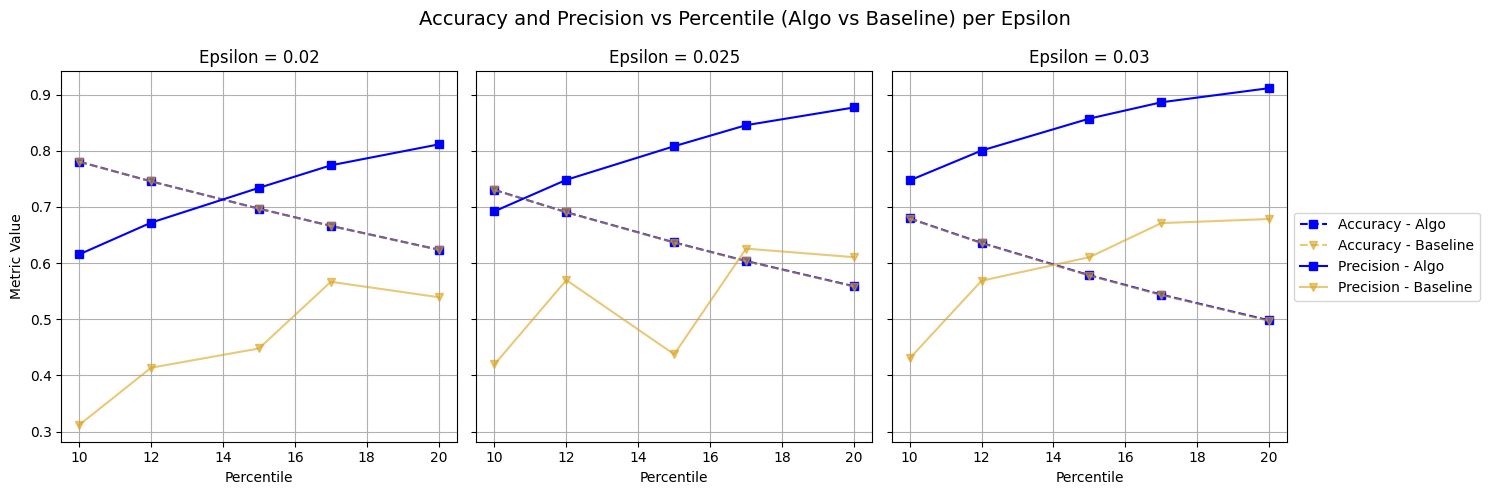}
    \caption{Classifier vs. baseline average accuracy and precision for Physics (arXiv) predictions 2010-2024 over a range of $\epsilon$ and $top-P\%$ cutoffs.}
    \label{fig:accuracyprecision} 
\end{figure}

Work is ongoing to optimize algorithm parameters for the best accuracy and precision tradeoffs for various applications. With some parameter settings, the algorithm's accuracy is barely distinguishable from baseline, but precision is notably higher. This difference persists across a range of $\epsilon$ and $top-P$, as shown in Figure~\ref{fig:accuracyprecision}. Such a result is to be expected for a dataset as unbalanced as ours: since we are classifying the $top-P\%$ articles where $P\approx15$, by definition a large majority of test set data points are negatives.

Possible applications of an algorithm with these characteristics include scenarios where false positives are particularly costly. One specific example could be investment allocation, measured in terms of the time and money which would be lost if dedicated to a false positive. In a competitive market with many opportunities, false negatives might be less costly than false positives.

\subsection{Reverse Engineering Prediction Coordinates}

Finding a spatial description of where valuable discoveries might be made in latent space is a useful and intriguing step, but it is not sufficient to contribute to the solution of the information overload alluded to at the beginning of this article. An important second step is the ability to translate the algorithmically predicted vectors into human-readable terms. This can be achieved using a method of reverse-engineering vectors back into natural language text, using an encoder-decoder model called Vec2Text~\citep{morris2023textembeddingsrevealalmost}. Vec2Text is more a proof-of-concept than a mature method, since it is limited to very short texts and sometimes produces nonsense results. However, it often works well, and demonstrates that it is possible to translate the predicted coordinates into actionable insight. Work is ongoing to improve the quality of the text output.

\section{Conclusion}
We conclude with a metaphor. The discovery of cosmic background radiation has led over the last six decades to successively more detailed maps of the universe. These maps exhibit an extremely faint but clearly measurable structure, and have spurred many researchers to study the connection between the early events in the evolution of the universe and the presently observable background radiation. The maps produced by our algorithm are similarly tantalizing: they bear witness to a faint but detectable structure within the distribution of scientific articles. Our further work will leverage the predictive value of that structure. 

\bibliography{references}

\end{document}